# Widely tunable dual acousto-optic interferometric device based on a hollow core fiber

[1,4]*Ricardo E. da Silva, [2]Jonas H. Osório, [3]Frédéric Gérôme, [3]Fetah Benabid,
[4]David J. Webb, [5]Marcos A. R. Franco, and [1]Cristiano M. B. Cordeiro

[1]*Institute of Physics Gleb Wataghin, University of Campinas (UNICAMP), Campinas, 13083-859, Brazil*
[2]*Department of Physics, Federal University of Lavras (UFLA), Lavras, 37200-900, Brazil*
[3]*GPPMM Group, XLIM Institute, UMR CNRS 7252, University of Limoges, Limoges, 87060, France*
[4]*Aston Institute of Photonic Technologies (AIPT), Aston University, Birmingham, B4 7ET, UK*
[5]*Institute for Advanced Studies (IEAv), São José dos Campos, 12228-001, Brazil*

ABSTRACT

An all-fiber dual Mach-Zehnder interferometer (MZI) based on an acoustically modulated hollow-core fiber (HCF) is experimentally demonstrated for the first time. By attaching an acoustic driver in between the fixed ends of an HCF, we fabricated two acousto-optic modulators (AOMs) with distinct driver positions, allowing for synchronizing two in-line MZIs inside the HCF. The first MZI is set by two acoustic long-period gratings separated by a second MZI formed at the fiber and driver attaching region. We show that this setup enables frequency-tuning of the coupling between the fundamental and higher-order modes in the HCF. Additionally, we simulate and analyze the HCF modal couplings and MZIs' modulated spectra under distinct device parameters using the transfer matrix method. The new AOM-MZI enables tuning of the MZIs free spectral range by adjusting 1 Hz of the electrical frequency, which is promising to modulate multiwavelength filters, sensors and fiber lasers.

## 1. Introduction

All-fiber Mach-Zehnder interferometers (MZI) have successfully been applied in mode converters [1,2], fiber sensing (fuel, gas, refractive index, temperature, curvature, strain, humidity [3–11]), optical vortex beam generation [2], spectral gain flattening of fiber amplifiers [12], and multiwavelength fiber lasers [13–18]. In these devices, the input optical beam is usually split by a coupler into two beams propagating through distinct optical paths (arms). Further, these beams are recombined by a second coupler, and an interference pattern is seen at the fiber output. The free spectral range (FSR) of the resulting interference fringes can be tuned by changing the beams' relative phase, e.g., by modifying the optical path length of one of the arms. With optical fibers, added flexibility can be gained by exploiting the multiple optical path lengths associated with distinct guided modes in the fiber. Additionally, distinct optical path lengths can be achieved in fiber MZIs by using concatenated-spliced fibers with special geometries (photonic crystal fibers [10,19], multicore fibers [20], suspended core fibers [9] and hollow core fibers (HCFs) [3]).

* Corresponding author at: r.da-silva@aston.ac.uk

The input and output MZI couplers can be fabricated by having two localized geometric or refractive index changes over the fiber cross-section, achieved by tapering [8,11] or micro-bending [7]. Alternatively, MZIs using long-period gratings (LPGs) as couplers add spectral filtering tunability and can increase the sensitivity to external measurands [21]. The LPG notch wavelength, bandwidth, and magnitude can be tuned by adjusting the grating period and modulated refractive index amplitude. Thus, MZIs employing a pair of permanently inscribed or mechanically induced LPGs have been employed in sensing and laser applications [1,2,4–6,12].

Acoustically induced LPGs couple power of the fundamental and higher order modes, enabling tuning of the modulated notch wavelength and depth by adjusting the acoustic frequency and amplitude [22,23]. The acousto-optic modulators (AOMs) usually employ tapered and etched single-mode standard fibers (SMFs), long fiber lengths, or high voltages (e.g., 38 V [24]) to increase the modulation efficiency. AOM-MZIs employing tapered fibers [16,25], etched fibers [15,17], etched fiber Bragg gratings (FBGs) [18] and two acoustic drivers have been reported [24,26]. In general, FSR tuning is usually limited to small spectral phase variations induced by applied temperature or strain. Hence, generation of multiwavelength fringes usually requires long fibers in MZIs, enabling FSR tuning by only replacing components or fabricating extra devices (e.g., MZI fiber lengths from about 9 to 20 cm should be replaced in AOMs to generate 9 to 19 wavelength fringes [15,25]). However, long fibers and additional components might increase the overall device size, consumed electric energy, response time, losses, and cost [24,26,27].

We have recently demonstrated highly efficient acoustically induced LPGs in tubular- and hybrid-lattice HCFs [22,23,28]. Here, we experimentally demonstrate a new all-fiber AOM design comprising two in-line MZIs inside an HCF displaying a hybrid kagome-tubular cladding. The demonstrated AOMs (with centered and off-center driver positions) provide unique features to dynamically generate a multiwavelength filter by frequency-tuning the FSR in steps of 1 Hz. In addition, the modulated spectra are also simulated and studied in detail considering different MZI parameters using the transfer matrix method (TMM).

## 2. Operation principle of the acousto-optic dual Mach-Zehnder interferometer

Acoustically induced LPGs (ALPGs) modulate the refractive index over the fiber cross-section. This modulation is mainly caused by bend-induced fiber geometric deformation changing the optical path length of the guided modes. Thus, ALPGs efficiently couple the power of a mode $LP_{KI}$ and a higher order mode $LP_{LJ}$ at the notch wavelength $\lambda_C$ when the modes beatlength $L_B$ phase matches with the acoustic period $\Lambda$ ($L_B = \Lambda$) [16–18,22], according to,

$$\lambda_C = (n_{KI} - n_{LJ})\Lambda, \qquad (1)$$

where,

$$L_B = \frac{\lambda_C}{n_{KI} - n_{LJ}}, \qquad (2)$$

$$\Lambda = \left(\frac{\pi r c_{ext}}{f}\right)^{\frac{1}{2}}, \quad (3)$$

$n_{KI}$ and $n_{LJ}$ are the effective refractive indices of $LP_{KI}$ and $LP_{LJ}$ ($K, L,$ are the mode azimuthal numbers and, $I, J$, the radial numbers), $r$ is the fiber radius, and $c_{ext}$ is the silica extensional acoustic velocity [29]. In this scenario, bend-induced axial strain changing the fiber optical path length might couple any modes having adjacent azimuthal numbers ($|K - L| = 1$), proper spatial overlap, and phase-matching as in (1) [30,31].

Fig. 1(a) illustrates the configuration we study herein, which encompasses an HCF segment fixed at its ends and an acoustic driver centrally fixed along a fiber contact length $d$. The acoustic driver excites simultaneously two identical ALPGs (induced by standing flexural acoustic waves of amplitude $A$, period $\Lambda$, and frequency $f$). We identify this configuration as AOM – MZI 1. ALPG 1 partially couples the input power ideally confined in the fundamental mode $LP_{01}$ to the first higher-order mode $LP_{11}$ modulating a notch in the HCF transmission spectrum (red line in Fig. 1(b)). The notch modulation depth and wavelength $\lambda_C$ can be tuned by adjusting the acoustic amplitude $A$ and frequency $f$, respectively. $LP_{01}$ and $LP_{11}$ propagate through distinct optical path lengths along the fiber experiencing a phase-shift as [5,6],

$$\phi \approx \frac{2\pi \Delta n_{eff} L_{CV}}{\lambda}, \quad (4)$$

where $\Delta n_{eff}$ is the modes' effective refractive index difference, $\lambda$ is the optical wavelength and, $L_{CV}$ is the ALPGs center-to-center separation (cavity length) [5,6,13,14]. ALPG 2 further recouples $LP_{01}$ and $LP_{11}$ causing them to interfere at the fiber output. The interference fringes' FSR is given as [13],

$$\text{FSR} \approx \frac{\lambda^2}{\Delta n_{eff} L_{CV}}. \quad (5)$$

Fig. 1(b) illustrates an arbitrary AOM - MZI interference spectrum caused by the coupling $LP_{01}$ - $LP_{11}$ (black solid line). In Fig. 1(a), we note that having $d$ tending to zero ($d \rightarrow 0$) allows matching the phase of ALPGs 1 and 2 resulting in only one acoustic grating along the HCF. This might also occur by considerably increasing the acoustic period $\Lambda$ compared to $d$ ($\Lambda \gg d$), indicating that $d$ should be longer than $\Lambda$ to produce proper gratings separation to generate multiple fringes. Alternatively, suitably choosing $d$ and a threshold frequency $f_0$, making $d$ smaller and comparable to $\Lambda$ (e.g., $d = \Lambda/2$) and neglecting $\Delta n_{eff}$ changes along $d$, the two gratings might work as a single ALPG with a local phase-shift [1,12].

We have recently investigated flexural acoustic waves in tubular HCFs with the 3D finite element method [23]. Current studies indicate that the pressure caused by ALPGs in the fixed fiber $d$ edges induces two strong curvatures, forming a second interferometer (MZI 2), as illustrated in Fig. 1(c). For fixed HCF ends, tuning the acoustic period with frequency $\Lambda(f)$ also tunes the critical curvature radius $R(f)$. Thus, MZI 2 similarly works as a two-point micro-bend MZI coupling higher order modes in addition to those coupled by the ALPGs, significantly increasing $n_{eff}$ by slightly decreasing the critical radius $R$ [7],

$$n_{eff}(R) = n\left(1 + \frac{u}{R}\right), \quad (6)$$

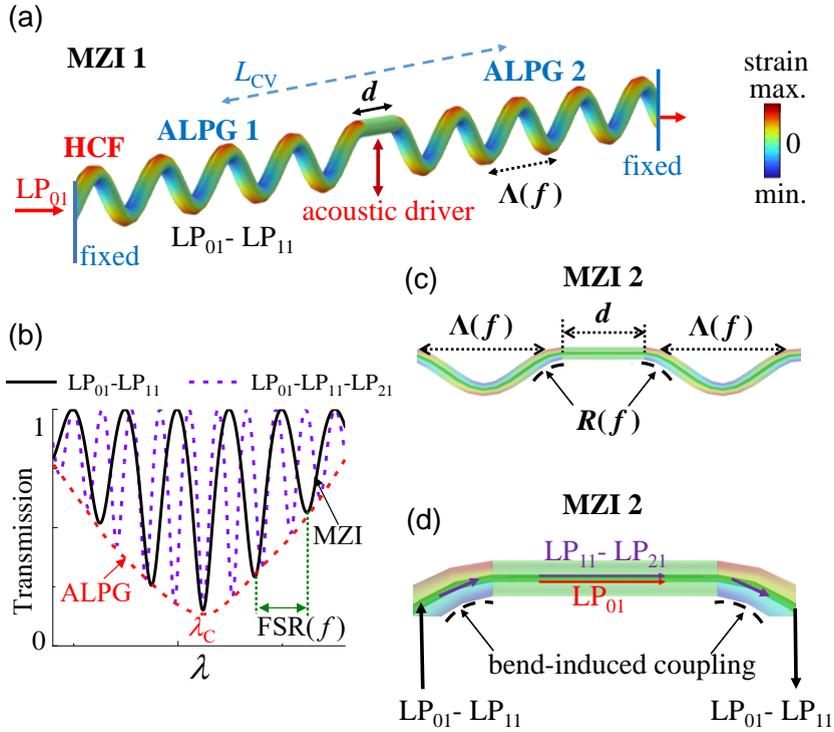

**Fig. 1.** (a) Illustration of a Mach–Zehnder interferometer (MZI 1) composed of two acoustically induced long period gratings (ALPG 1 and ALPG 2), excited at the central fiber length $d$, coupling the fundamental $LP_{01}$ and higher order mode $LP_{11}$ in a hollow core fiber (HCF). (b) HCF transmission spectrum tuning the fringes' free spectral range (FSR) by distinct bend-induced modal couplings in MZI 2. (c) The critical curvature radius $R$ is changed by tuning the acoustic wave period with frequency $\Lambda(f)$. (d) MZI 2 couples higher order modes ($LP_{11}$ and $LP_{21}$) increasing $\Delta n_{eff}$ and the number of fringes in the spectrum in Fig. 1(b) (purple dashed line).

where, $n$ is the unbent fiber refractive index and, $u$ is the orientation vector from the bent fiber center. Accordingly, increasing $\Delta n_{eff}$ reduces the fringes' FSR in (5) contributing to increase the number of fringes in the spectrum. For example, Fig. 1(d) illustrates the coupling of $LP_{11}$ (coming from ALPG 1) and the higher order mode $LP_{21}$, increasing the overall difference of optical path length with $LP_{01}$. The coupling between the modes in the dual MZI follows the coupling sequence, $LP_{01} \to LP_{11} \to LP_{21}$ [30,31]. The resulting increased $\Delta n_{eff}$ modulates new fringes in the HCF spectrum, as shown in Fig. 1(b) (purple dashed line). We note that MZI 2 therefore provides a frequency-tunable multipath arm for MZI 1.

We start by analyzing the spectral characteristics of the HCF MZI and the corresponding dependences on its construction parameters by simulating it using the TMM (a detailed description of the employed formulation and modeling methods is provided in [12,19,21]). The ALPGs parameters used in the simulations are based on the experimental components and measured modulated spectra described in the next section. Fig. 2(a) shows the HCF MZI simulated transmission spectrum indicating the acoustically modulated notches for changing $d$ from 1.54 to 1.64 mm ($\Delta d = 100$ µm). Here, we set the acoustic driver centered at the fiber length $L$ ($P = 0.5L$) and $\Delta n_{eff} = 2.8 \times 10^{-3}$ (in this configuration, MZI 2 modal couplings inducing multiple fringes are negligible so that $L_{CV} = d$). Note that $d = 1.54$ mm induces nearly a $\pi$-phase-shift generating two notches in the spectrum (which might also be achieved by having minor $n_{eff}$ changes in MZI 2). In contrast, only one notch is modulated when $d = 1.64$ mm ($\phi = 2\pi$).

Fig. 2(b) shows the results for the situation in which the driver has been further shifted to $P = 0.25L$ and $0.75L$ and the modulated spectrum evaluated at $\phi = 1\pi$ and $\phi = 2\pi$. Note that changing $P$ is irrelevant at $\phi = 2\pi$ while the fringes visibility decreases for the off-center driver positions at $\phi = 1\pi$ (no difference is observed for any off-center $P$).

In turn, Fig. 2(c) shows the modulated spectra for increasing $\Delta n_{eff}$ by about twice (black line) and 4 times (blue line) in MZI 2. As expected, decreasing FSR almost doubles the number of fringes from $N_F \sim 7$ to 14. For a fixed $\Delta n_{eff}$, the off-center driver would similarly reduce the fringes visibility regardless of $\phi$, as seen in Fig. 2(d) and 2(e) (fringe visibility is the effective fringe magnitude or height). Overall, properly selecting $d$ to adjust the phase shift might be useful to equalize the magnitude of the central fringes.

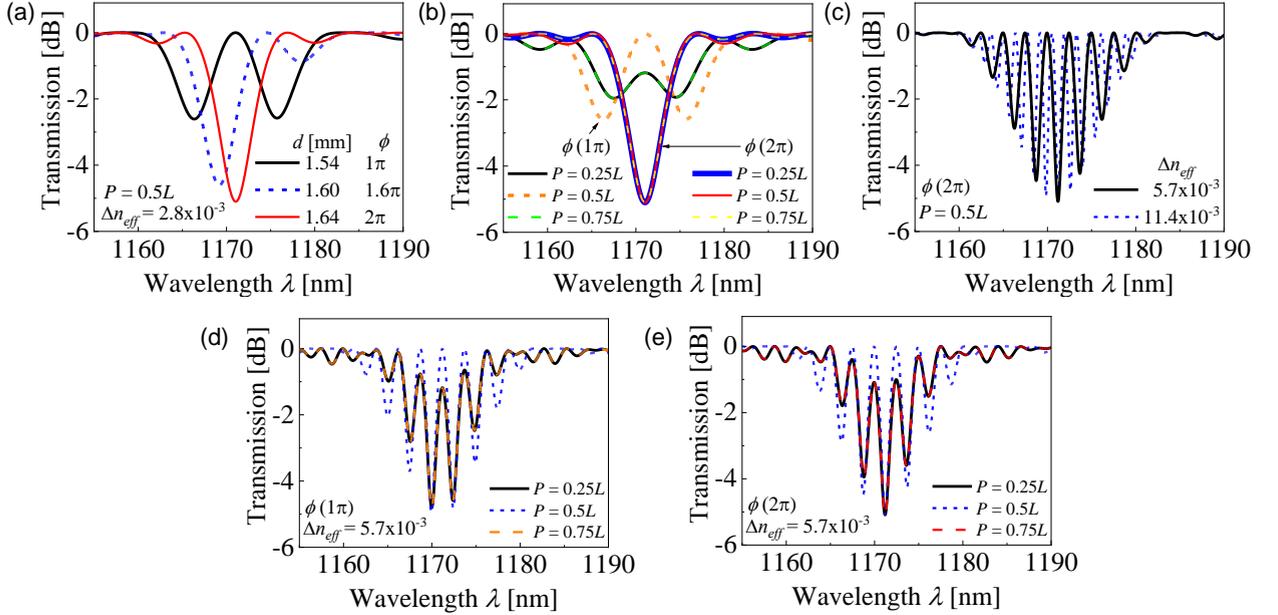

**Fig. 2.** HCF acoustically modulated spectra simulated with the transfer matrix method (TMM). (a) MZI 1 phase-shift variations caused by distinct fixed fiber lengths $d$ and (b) by shifting the acoustic driver along the fiber length $L$ at the positions $P = 0.25L$, $P = 0.5L$ and $P = 0.75L$ (for $\phi = 1\ \pi$ and $2\pi$ phase-shifts in Fig. 2(a)). (c) The increasing difference of the modes effective refractive index $\Delta n_{eff}$ adds wavelength fringes with decreased spectral separation. The fringes visibility is reduced for off-center acoustic excitation regardless the initial phase-shift $\phi$ of (d) $1\pi$ or (e) $2\pi$.

## 3. Experimental setup

The fabricated AOMs employ a hybrid cladding HCF composed of a Kagomé structure (800 nm struts thickness) and six suspended tubes (20 µm diameter and 1.25 µm thickness) forming the air core (34 µm diameter), as shown in Fig. 3(a). The HCF is made of silica and has a 240 µm outer diameter. Additional details about the HCF design, fabrication, and modal characterization are described in [32].

Two AOMs are assembled to study the HCF modulated transmission spectrum. Fig. 3(b) illustrates the AOM 1 setup, composed of an HCF segment of length $L$, a piezoelectric transducer disc (PZT), and a transversely connected silica acoustic horn. PZT is 2 mm thick and 25 mm in diameter. The horn is tapered in diameter from 4.8 to 0.65 mm along 18 mm. PZT, horn, and connections are similar to those in [22]. The HCF ends are fixed forming an interaction length of $L = 75$ mm. The HCF input is butt-coupled to a single mode fiber (SMF) that is lens-coupled to a supercontinuum laser source (SC). The HCF output is butt-coupled to a multimode fiber (MMF) connected to an optical spectrum analyzer (OSA). The coupling alignments are performed using micrometer stages and mirrors.

Fig. 3(c) illustrates details of the dual MZI, indicating the connection of the horn and HCF. The fiber is shifted from the horn tip to increase the fiber contact length $d$ with the horn. The horn is centered at the position $P = 0.5L$ and fixed to the HCF with adhesive along $d \sim 1.6$ mm (red circle in Fig. 3(c)). Thus, the HCF is split into two ALPGs with equivalent lengths ($L_1 \sim L_2 \sim 36.7$ mm), forming an effective MZI 1 cavity length of $L_{CV} \sim 38.3$ mm. The separation $d$ between the gratings corresponds approximately to MZI 2. A sinusoidal electrical signal is applied to PZT using a signal generator (SG) with a maximum 10 V voltage (no amplifier has been used). A HCF transmission spectrum notch at $\lambda_C = 1170$ nm is tuned at $f_0 = 263.029$ kHz. The fringes' FSR is tuned by adjusting frequency from $f = 263.027$ to $263.031$ kHz (1 Hz steps).

As mentioned above, we assembled a second AOM (AOM 2) employing the same components as in AOM 1. In AOM2, however, the acoustic horn is 20 mm shifted from the HCF right-end ($P \sim 0.25L$), and ALPGs lengths are changed accordingly ($L_2 \sim 19$ mm and $L_1 \sim 54$ mm; $d$ and $L_{CV}$ remains with similar dimensions as in AOM 1). The modulated notch at $\lambda_C = 893$ nm is set at $f_0 = 269.329$ kHz and the fringes' FSR is tuned in the range of $f = 269.327 - 269.331$ kHz (1 Hz step). Although not shown in this manuscript, other modulated notches could also be tuned in the range of $\lambda = 800 – 1200$ nm. The considered spectral notches have been chosen due to their interest for multiwavelength lasers, particularly around the gain range of Ytterbium-doped fiber lasers [33]. The measured and simulated spectra of AOMs 1 and 2 are compared and analyzed in detail respectively in Sec. IV-A and IV-B.

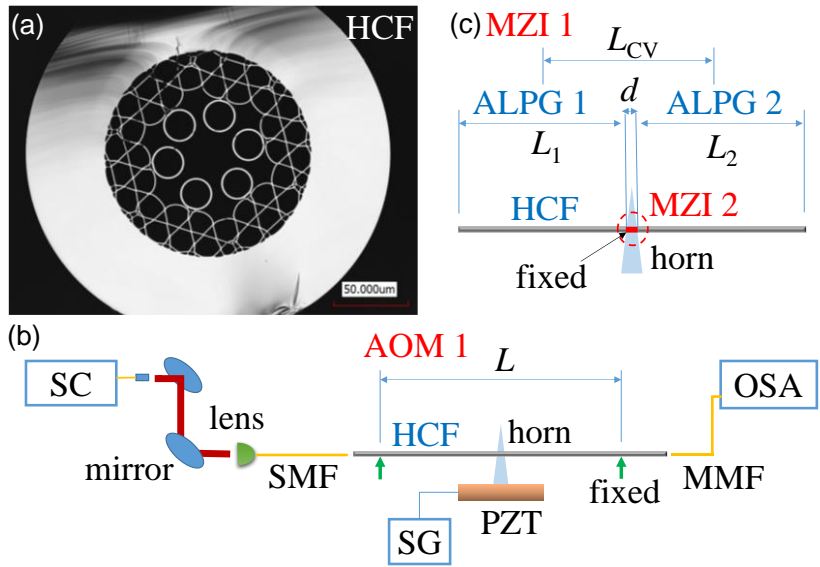

**Fig. 3.** (a) Hybrid cladding hollow core fiber (HCF). (b) Experimental setup and acousto-optic modulator (AOM) using a piezoelectric transducer (PZT), an acoustic horn and an HCF segment of length $L$ (horn is placed at AOM 1 center (0.5$L$) and further right-shifted in AOM 2 (0.25$L$ – not shown). (c) Details of the two Mach-Zehnder interferometers (MZIs) in AOM 1.

## 4. Results and discussion

4.1 FSR tuning of the multiwavelength filter with centered acoustic driver (AOM 1)

We have measured and simulated AOM 1 transmission spectra with the centered acoustic driver. Fig. 4(a) shows the simulated HCF modes' beatlengths $L_B$ compared to the acoustic period $\Lambda$ calculated from the measured frequency $f = 263.029$ kHz ($L_B$ and $\Lambda$ are calculated for the applied frequencies using (1) - (3) [23]). The measured notch wavelength $\lambda_C = 1170$ nm is indicated with a vertical dashed line in Fig. 4(a). The HCF modes' effective refractive indices $n_{eff}$ are computed employing the actual HCF dimensions, and analytical models described in [34] ($n_{eff}$ is computed for an arbitrary mode $LP_{KI}$ sweeping the indices $K$ and $I$ from 1 to 5). Note that $\Lambda$ matches simultaneously with multiple $L_B$, indicating ALPGs couplings of the fundamental $LP_{01}$ and higher order modes $LP_{11}, LP_{21}, LP_{31}, LP_{32}, LP_{41},$ and $LP_{51}$ (this multi-coupling is caused by the modes $\Delta n_{eff}$ and $L_B$ nearly matching when $\lambda_C$ approaches the edge of the HCF's transmission band [32]).

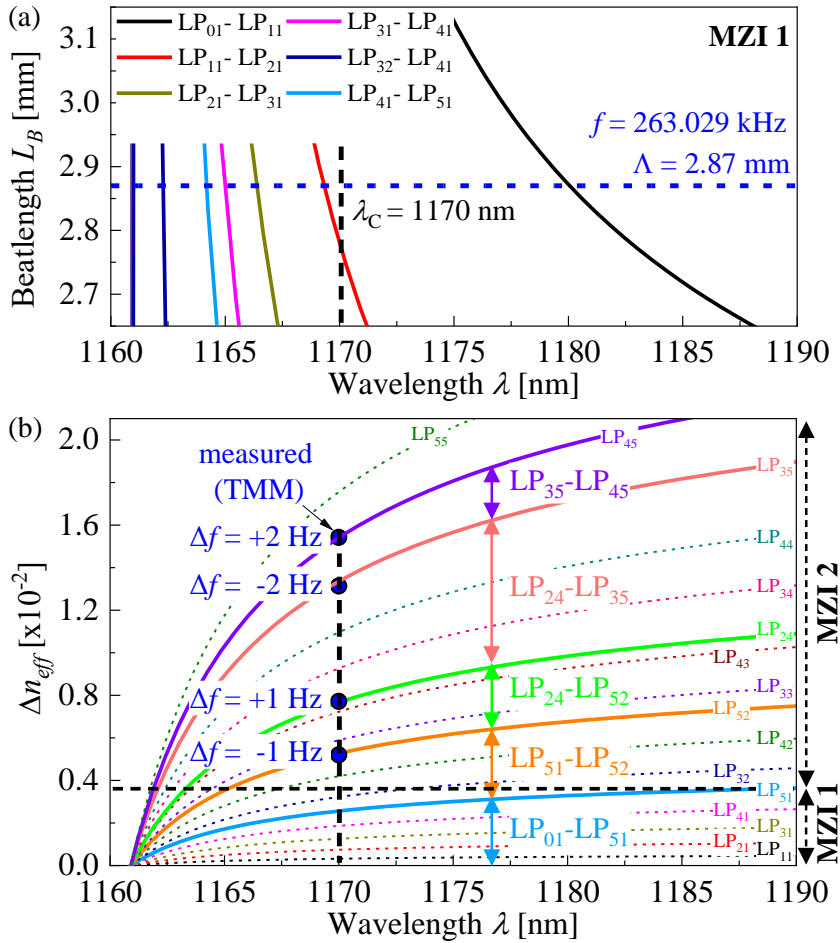

**Fig. 4.** (a) AOM 1 - ALPGs phase matching in MZI 1 showing the simulated HCF core modes' beatlengths $L_B$ matching the acoustic period $\Lambda$ over the measured notch bandwidth around $\lambda_C$ (dashed black line). (b) Effective refractive index difference $\Delta n_{eff}$ of the fundamental $LP_{01}$ and relevant higher order modes coupling in MZI 1 and MZI 2 (main couplings are indicated with arrows and solid lines). The simulations are compared to $\Delta n_{eff}$ estimated from the measured-simulated spectra for tuning frequency of $\Delta f = \pm 1$ and $\pm 2$ Hz in Fig. 5 (blue circles).

Fig. 4(b) shows the modes $\Delta n_{eff}$ calculated with respect to $LP_{01}$. Thus, MZI 1 couples favorably in sequence modes from $LP_{01}$ to $LP_{51}$ ($LP_{01}$-$LP_{51}$) as discussed in Sec. II. Here, only core modes having relevant spatial overlap of electric fields over the HCF cross-section are considered (details are described in [23]). Still, other modes with low spatial overlap (e.g., tubular modes) might also couple in MZI 1 with negligible modulation efficiency. These modes are therefore ignored in the simulations since they do not affect our analyses. Fig. 4(b) shows the $\Delta n_{eff}$ of the modal couplings in MZI 2 (indicated by arrows and solid lines). Additionally, Fig. 4(b) displays the $\Delta n_{eff}$ estimated from the measured and simulated spectra in Fig. 5 at frequency steps of $\Delta f = \pm 1$ and $\pm 2$ Hz (blue circles). We observe good agreement between the $\Delta n_{eff}$ values and note that FSR tuning is therefore defined by frequency-tuning MZI 2 edges' curvatures coupling higher order modes.

The measured and simulated spectra of AOM 1 for tuning frequency from $f = 263.027$ to $263.031$ kHz are thus shown in Fig. 5. Fig. 5(a) shows a nearly π-phase-shift modulated spectrum indicating operation of MZI 1 at the frequency threshold of $f_0 = 263.029$ kHz. Fig. 5(a) indicates the ALPGs forming a unique grating with a π-phase-shift induced by minor changes of the experimental fixed fiber length $d$ ($d = 1.54$ mm is estimated from TMM). The modulated spectrum might also have minor $\Delta n_{eff}$ changes induced in MZI 2. In contrast, a multiwavelength spectrum is achieved by slightly decreasing $f$ to 263.028 kHz ($\Delta f = -1$ Hz), as shown in Fig. 5(b). Agreement of measured and simulated spectra implies synchronization of MZIs 1 and 2 (the flexural acoustic waves have the same polarization along the HCF in both MZIs). The measured spectrum shows high fringe visibility in Fig. 5(b).

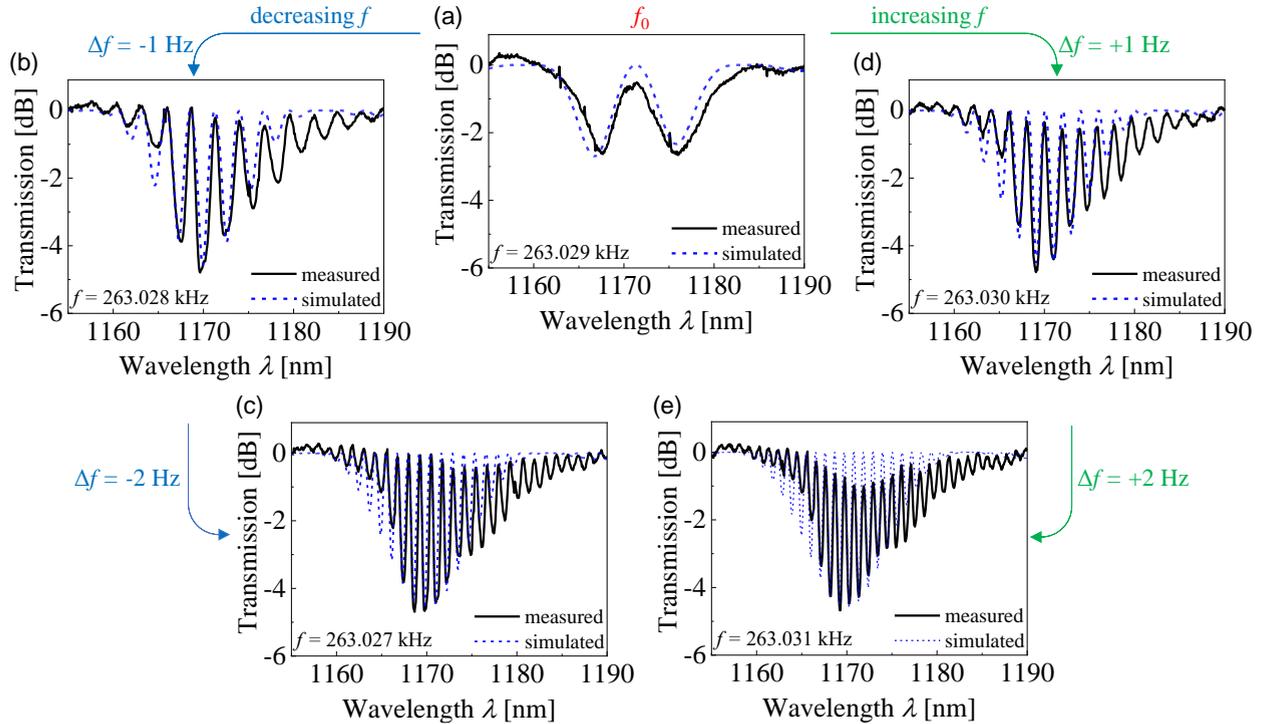

**Fig. 5.** AOM 1 measured and simulated transmission spectra of the hollow core fiber for the frequency range of $f = 263.027 - 263.031$ kHz. (a) MZI 1 induces a nearly π-phase-shift in the modulated spectrum at the threshold frequency $f_0$. Dual MZI multiwavelength spectra by tuning the fringes' FSR with (b)(c) decreasing frequency and (d)(e) increasing frequency respectively in steps of $\Delta f = 1$ and 2 Hz.

In turn, the fringes' FSR is significantly reduced by decreasing 2 Hz, inducing new fringes in the notch (Fig. 5(c)). This highly sensitive FSR tuning indicates strong bend-induced modal couplings in MZI 2 (Fig. 4(b)). Similarly, FSR tuning is achieved by finely increasing frequency at $f$ = 263.030 and 263.031 kHz, as shown in Fig. 5(d) and 5(e) (fringes parameters for increasing $f(+)$ and decreasing frequency $f(-)$ are further summarized in Fig. 8, denoting slightly distinct FSR and $\Delta n_{eff}$ in these cases). Moreover, reduced fringe visibility in Fig. 5(e) implies that increasing modal couplings might induce power losses (max. 1 dB). Overall, the average fringe FSR could be tuned from FSR = 8 to 1 nm with a tuning slope of 0.9 nm/Hz (inducing 9 fringes/Hz), as shown in Fig. 8(a) and 8(b) ($f_0$ spectrum is not included in FSR estimate).

The modulated spectra show a 3-dB notch bandwidth of 8 nm and an average maximum modulation depth of 4.7 dB at about $\lambda_C$ = 1170 nm for the considered frequencies, excepting $f_0$. The notch shows asymmetric fringe distribution around $\lambda_C$, deviating from an ideal Gaussian-like profile, suggesting predominant coupling of $LP_{11}$ and $LP_{21}$ with high field overlap [23,32] (note that the measured spectra in Fig. 5 follow closely the simulated asymmetric modal distribution shown in Fig. 4(a)).

## 4.2 FSR tuning of the multiwavelength filter with off-center acoustic driver (AOM 2)

The measured and simulated transmission spectra for the configuration displaying an off-center acoustic driver (AOM 2) are shown in Fig. 6. The modulated spectra are tuned in the frequency range of $f$ = 269.327 - 269.331 kHz. Fig. 6(a) shows a nearly $2\pi$-phase-shift notch indicating operation of MZI 1 at $f_0$ = 269.329 kHz. In contrast to AOM 1, only one notch is observed ($d$ and $\Delta n_{eff}$ minor changes contribute to distinct phase shifts at $f_0$).

A multiwavelength spectrum is generated by tuning at $f$ = 269.328 kHz ($\Delta f$ = -1 Hz), as seen in Fig. 6(b). Measured and simulated fringes showing decreased fringe visibility (magnitude variation of less than 1 dB) indicate the effects of the off-center driver position, as expected from the simulations (Fig. 2(e)). The fringes FSR is further reduced by decreasing $\Delta f$ = 2 Hz, inducing new interference fringes in the notch (Fig. 6(c)). The measured high fringe visibility indicates that increasing modal couplings in MZI 2 might induce a stronger effect on fringes visibility compared to the off-center driver position (it is also noted in AOM 1). Overall, the fringe FSR is also tuned by increasing frequency, as shown in Fig. 6(d) and 6(e) (note that the fringe number $N_F$ nearly doubles at $\Delta f$ = +2 Hz compared to $N_F$ at $\Delta f$ = +1 Hz). The fringes are tuned from FSR = 13 to 6 nm (average) with a slope of 6 nm/Hz (4 fringes/Hz). The tuned fringes' parameters are further summarized in Fig. 8, indicating similar FSR, $N_F$, and $\Delta n_{eff}$ for decreasing and increasing $f$.

Fig. 6 shows that the fringe distribution in the notches is nearly symmetric around $\lambda_C$ = 893 nm, resembling a Gaussian-like profile with minor phase variations. As expected, improved notch symmetry is caused by reduced modal couplings along the ALPGs, preferably coupling $LP_{01}$ and $LP_{11}$, as indicated in Fig. 7(a) ($\lambda_C$ closely matches $L_B - \Lambda$). This decreased number of couplings compared to AOM 1 is due to the larger difference between the modes $\Delta n_{ef}$ as $\lambda_C$ approaches the center of the HCF's transmission band [32]. In addition, $\Delta n_{eff}$ changes slightly over the spectrum also contributing to notch symmetry. This is also noted for modal couplings in MZI 2 in Fig. 7(b). The frequency-increased $\Delta n_{eff}$ (blue circles) might be caused by couplings $LP_{11}$ - $LP_{21}$ ($\Delta f = \pm 1$ Hz) and $LP_{12}$ - $LP_{21}$ ($\Delta f = \pm 2$ Hz) (indicated with arrows and solid lines).

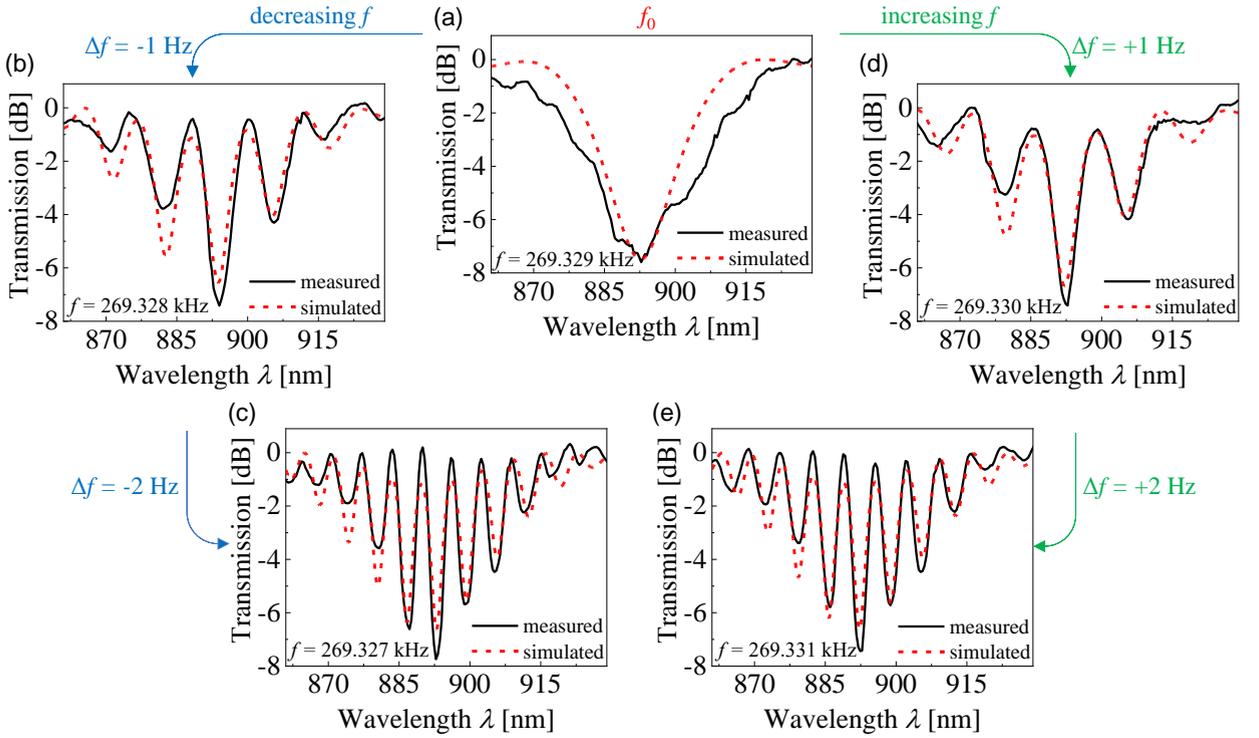

**Fig. 6.** AOM 2 measured and simulated transmission spectra of the hollow core fiber for the frequency range of $f$ = 269.327 - 269.331 kHz. (a) MZI 1 induces a nearly $2\pi$-phase-shift in the modulated spectrum at the threshold frequency $f_0$. Dual MZI multiwavelength spectra by tuning the fringes FSR with (b)(c) decreasing frequency and (d)(e) increasing frequency in steps of $\Delta f$ = 1 and 2 Hz.

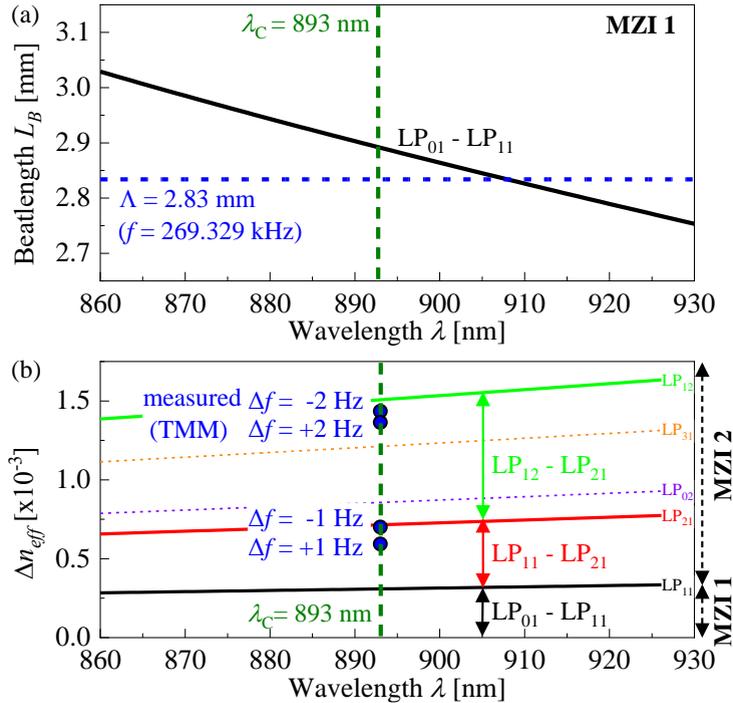

**Fig. 7.** (a) AOM 2 - ALPGs phase matching in MZI 1 showing the beatlength $L_B$ of the fundamental $LP_{01}$ and first higher order mode $LP_{11}$ nearly matching the acoustic period $\Lambda$ over the measured notch bandwidth around the $\lambda_C$. (b) Effective refractive index difference $\Delta n_{eff}$ of the fundamental $LP_{01}$ and relevant higher order modes in the HCF core. The modal couplings in MZI 1 (black arrow) and MZI 2 (red and green arrows) are compared to $\Delta n_{eff}$ estimated from the measured-simulated spectra in Fig. 6 (blue circles).

The estimated measured-simulated $\Delta n_{eff}$ changing with frequency is shown in Fig. 8(c) (blue line). Overall, the modulated spectra show a relevant maximum modulation depth of 7.5 dB at $\lambda_C = 893$ nm. The notch 3-dB bandwidth changes from 24 to 28 nm (3x AOM 1 bandwidth). Thus, decreasing modal couplings, and reducing $\Delta n_{eff}$ in ALPGs might be useful to broaden the notch width at the expense of reduced modulated fringes, as indicated for AOM 1 and 2 in Fig. 8(b). Although it is not shown herein, we have observed increasing modulated fringes at frequencies higher than $\Delta f = 2$ Hz, suggesting that HCF modal properties might still provide broader $\Delta n_{eff}$ and FSR tuning range by coupling modes with near effective refractive indices (these modes usually have high losses and do not propagate long lengths in the unbent HCF [32]).

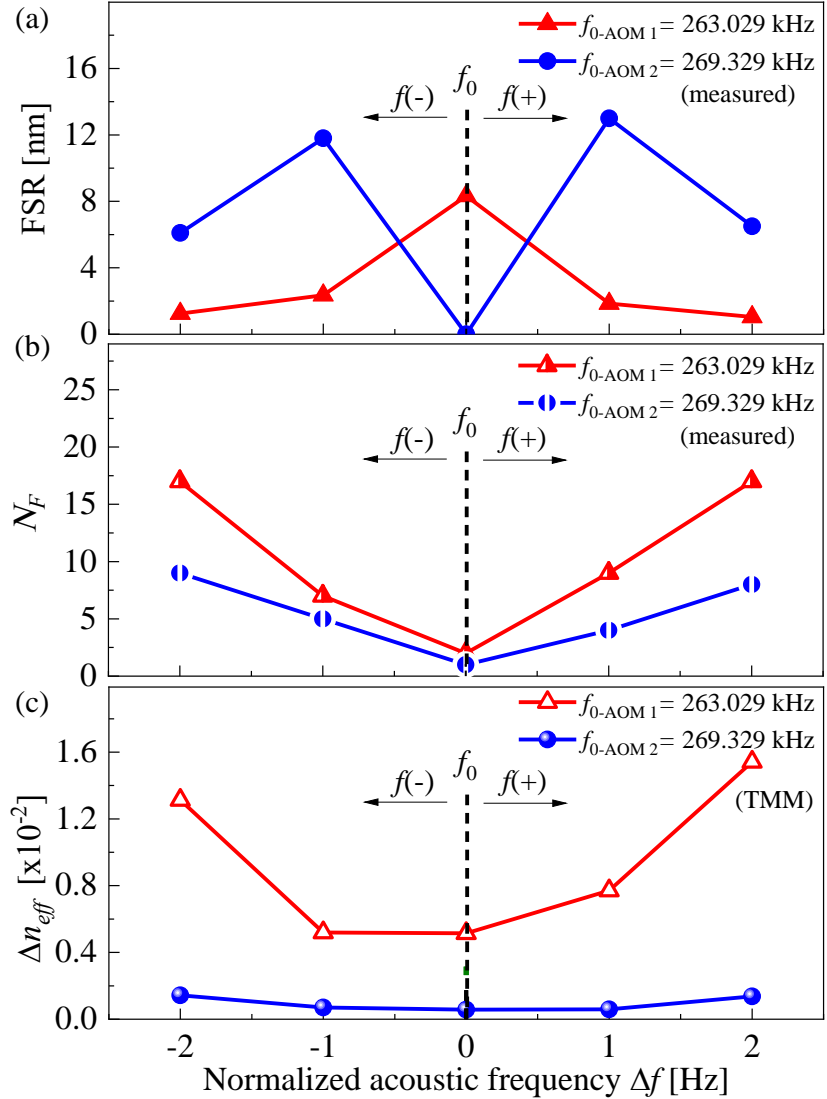

**Fig. 8.** (a) AOM tuning of the fringes' free spectral range (FSR) by decreasing $f(-)$ and increasing $f(+)$ the threshold frequency $f_0$ in 1 Hz steps (AOM 1 – red, AOM 2 – blue). Variation of the (b) fringe number $N_F$ and (c) modes' effective refractive index difference $\Delta n_{eff}$ for the considered frequency steps.

In summary, to the best of our knowledge, we demonstrate the first all-fiber acousto-optic dual MZI providing a fully reconfigurable multiwavelength filter with frequency-tunable FSR. Additionally, the HCF modal couplings and acoustically modulated spectra are evaluated in detail with TMM. The results indicate that changes in fringe visibility caused by changing the driver position (centered or off-center) are negligible compared to those induced by multimode coupling. Differences between measured and simulated spectra might be caused by unpredicted variations of the mode's effective indices along the HCF (e.g., caused by minor changes in the HCF geometry, dimensions, and material), as well as variations of acoustic parameters, such as ALPGs amplitude, period and length (and edge curvatures and fixed fiber length). Overall, the demonstrated AOMs show significantly high modulation efficiency employing low drive voltages (10 V) and a short fiber length (7.5 cm). This improved efficiency is attributed to increased acoustically induced axial strain in the HCF tubes changing the optical guidance of modes in the air core [28]. This enables the use of large diameter HCFs reducing requirements of tapering, etching, long fiber lengths, high voltages, power amplifiers, or strict connection of the fiber and acoustic horn (HCF might be vertically moved from the horn tip to adjust the fixed fiber length $d$). Improved modulation efficiency might still be achieved by using HCFs with larger air holes, thinner tubes, or reduced diameters, and optimizing the design of the acousto-optic device [23]. Further studies might also explore fibers with distinct geometries, tailoring the spectral modal properties to improve FSR sensitivity and modulated multiwavelength spectrum.

## 5. Conclusion

We have experimentally demonstrated an all-fiber acousto-optic dual Mach–Zehnder interferometer in an HCF for the first time. The two fabricated AOMs excite a pair of dynamic ALPGs (MZI 1) separated by an HCF segment transversally excited (MZI 2). The modulated multiwavelength spectra are compared to analytical simulations using the transfer matrix method. AOM 1 (centered driver position) modulates a notch bandwidth of 8 nm with a maximum depth of 4.7 dB at 1170 nm. FSR is tuned from 8 to 1 nm by changing frequency steps of 1 and 2 Hz (0.9 nm/Hz slope). AOM 2 (off-center driver position) modulates a 28 nm wide notch with a maximum of 7.5 dB at 893 nm (FSR = 13 – 6 nm with a 6 nm/Hz slope). Thus, AOM 1 provides higher frequency-tuning sensitivity inducing more fringes (17) in a narrower asymmetric notch, due to increased modal couplings and $\Delta n_{\text{eff}}$. In turn, AOM 2 shows a three times broader nearly symmetric notch with higher fringe visibility. Overall, measured-simulated spectra and modal analysis show a strong dependence of the modulated notch width, fringe number, and FSR on the induced HCF modal couplings changing $\Delta n_{\text{eff}}$. This is suitable for tailoring multiwavelength filters by adjusting the HCF geometry. The demonstrated compact 7.5 cm long AOMs provide high modulation efficiency employing fewer components, enabling promising tuning of reconfigurable fiber sensors and lasers.

## Declaration of Competing Interest

The authors declare that they have no known competing financial interests or personal relationships that could have appeared to influence the work reported in this paper.

**Acknowledgements**

This work was supported by the São Paulo Research Foundation (FAPESP) [grants #2022/10584-9, #2024/02995-4]; Conselho Nacional de Desenv. Científico e Tecnológico (CNPq) [grants 305024/2023-0, 305321/2023-4]; and Minas Gerais Research Foundation (FAPEMIG) [grants RED-00046-23, APQ-00197-24].